\documentclass{PoS}

\title{The interplay between cosmology, particle physics and astrophysics\thanks{Drawing largely but not exclusively on work presented in \cite{Vincent:2014rja} and \cite{Wilkinson:2016gsy}.}}

\ShortTitle{Cosmology, particle physics \& astrophysics}

\author{Aaron C. Vincent\\
        Arthur B. McDonald Canadian Astroparticle Physics Research Institute, Department of Physics, Engineering Physics and Astronomy, Queen's University, Kingston ON K7L 3N6, Canada\\Visiting Fellow, Perimeter Institute for Theoretical Physics, Waterloo ON N2L 2Y5, Canada
        E-mail: \email{aaron.vincent@queensu.ca}}

\abstract{As hints of physics beyond the standard model become the major driving force behind future large-scale projects, it is increasingly important to consider all sources of evidence and constraints. Here, I illustrate the importance of considering the connection between particle physics, cosmology and astrophysics, mainly via two examples: 1) the sterile neutrino and its impact on cosmology, and  2) the 511 keV line from electron-positron annihilation in the galactic centre.}

\FullConference{2nd World Summit: Exploring the Dark Side of the Universe\\
		25-29 June, 2018\\
		University of Antilles, Pointe-\`A-Pitre, Guadeloupe, France}

\begin{document}

\section{Introduction}
\begin{figure}
\begin{center}
\includegraphics[width=0.6\textwidth]{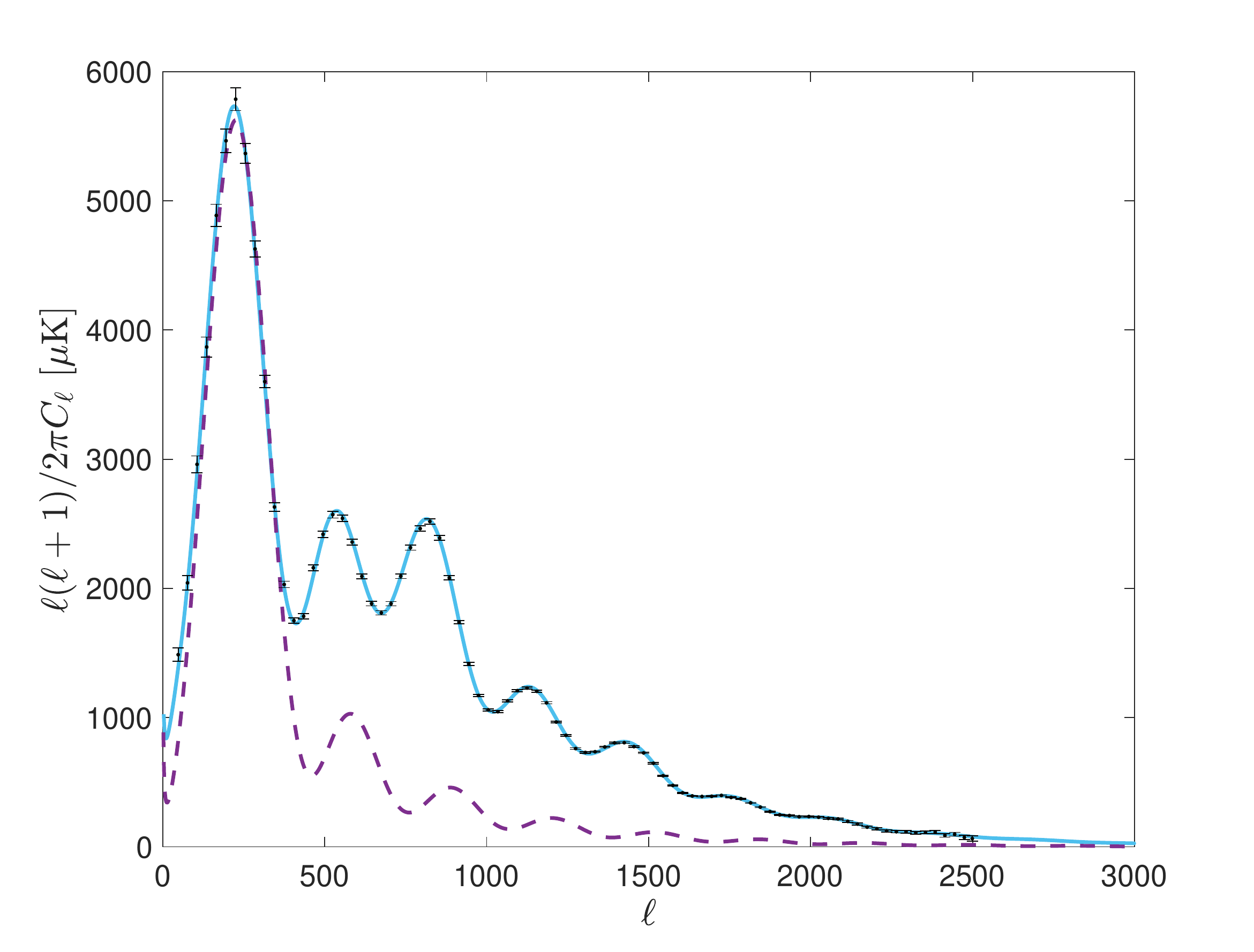}
\end{center}
\caption{The angular temperature anisotropy power spectrum of the cosmic microwave background with (top, blue) and without (bottom, purple) including a dark matter component. Other $\Lambda$CDM parameters have been tuned to obtain the first acoustic peak. Black points are data from Planck \cite{Ade:2015xua}. Nonbaryonic  CDM is collisionless and therefore unaffected by the large amounts of Silk damping seen in a baryonic universe.}
\label{fig:cmbdm}
\end{figure}

The Standard Model of Particle Physics has had tremendous success. All but four\footnote{The electron, photon, muon (``who ordered that?'') and tau. The latter was predicted on the basis that ``since muons exist in nature for no apparent reason, it is possible that other heavy leptons may also exist in nature'' \cite{PhysRevD.4.2821}.} of the SM particles were arguably predicted from the structure and known components of the SM itself. The fact remains that the SM itself is incomplete: SM neutrinos are massless, while their oscillation indicates nonzero mass splittings. Issues of hierarchy and tuning (such as the strong CP problem) hint at new physics at higher scales. It has long been known that extensions to the standard models have consequences far beyond the lab: for instance, new supersymmetric particle can affect everything from nucleosynthesis, to cosmological evolution, to galactic dynamics and stellar evolution. At the same time, a number of problems have arisen in cosmology and astrophysics that appear to require a modification either of the laws of physics on large scales, or a new nonstandard ingredient in the Universe. Galactic rotation curves, strong lensing and galaxy formation time scales all point to a large amount of nonbaryonic dark matter in the present Universe. Observations on large scales, from distant Type Ia supernovae to the scale of baryon acoustic oscillations (BAO) and the temperature and polarization anisotropies of the cosmic microwave background (CMB) all point to either a fundamental misunderstanding of cosmology and gravity, or a universe that is composed of 70\% dark energy, 25\% dark matter, with only about 5\% of ordinary baryonic matter and radiation. In Fig.~\ref{fig:cmbdm} I illustrate the simple yet striking observation that a Universe without cold dark matter produces a very different power spectrum from the one we observe: the baryonic pressure causes fluctuations on small scales to be severely damped with respect to the $\Lambda$CDM universe. 

Particle dark matter in particular offers many examples of clear connections between cosmology, particle physics and astrophysics. Many BSM models of particle physics predict new, cosmologically stable particles with no electric charge. The axion and the neutralino (or, more broadly the lightest supersymmetric partner) are canonical examples. Producing the correct relic abundance of dark matter to satisfy the cosmological dark matter problem is then a question of couplings and initial conditions, which may require more or less tuning within each model depending on the scenario. Of particular interest are thermally-produced models, in which the relic abundance is set by thermodynamics once the parameters of the theory (usually couplings and masses) are determined. Conversely, particle models can be quickly constrained and discarded by considering basic cosmological and astrophysical predictions of their effects.

In this short document I illustrate the interplay between cosmology, particle physics and astrophysics using two examples: in Sec.~\ref{sec:sterile} I describe the freeze-in production of a fourth-generation sterile neutrino that mixes with the standard model neutrinos; in Sec.~\ref{sec:511} I discuss the long-standing 511 keV excess from $e^+e^-$ annihilation near the Galactic centre, and its particle dark matter interpretation. I muse about what it all means in Sec. \ref{sec:conclusions}.

\section{Example 1: freeze-in of a heavy sterile neutrino}
\label{sec:sterile}
Extensions of the SM aiming to explain the masses of left-handed neutrinos generically introduce a new ``sterile'' neutrino state, which couples to its SM counterparts via some effective mixing angle $\sin\theta$. It will thus be produced in the early Universe via \textit{freeze-in}~\cite{Dodelson:1993je}. This involves the production of a thermal relic population of feebly interacting particles in the thermal bath of the early Universe, from an initially negligible number density. In Ref.~\cite{Vincent:2014rja} we considered the impact of such a new particle on cosmological observables. Such a particle is produced from mixing with active neutrinos, as described e.g. in \cite{Dolgov:2002wy}. Its number density evolution is then governed by Hubble expansion and by decay into lighter particles.

 The time-dependent evolution of the sterile and active species can be written in the simple way:
\begin{eqnarray}
  \Omega_ { s } ^ { \prime } ( x ) &=& - 3 \left( 1 + w _ { s } ( x ) \right) \Omega_ { s } ( x ) - \frac { \gamma _ { s } } { \mathcal { H } ( x ) } \Omega_ { s } ( x )  \\  \Omega_ { \nu } ^ { \prime } ( x ) &=& - 4 \Omega_ { \nu } ( x ) + \frac { \gamma _ { s } } { \mathcal { H } ( x ) } \Omega_ { s } ( x ) 
\end{eqnarray}
Here $x \equiv \ln a(t)$, is the log of the scale factor, $\Omega_i = \rho_i/\rho_c$. This allows for a dimensionless Friedmann equation $\mathcal{H}^2(x) = \sum_i y_i  \equiv H^2/(8\pi G \rho_c/3)$. The equation of state $w_{s}(x)$ is a function of the sterile neutrino mass and temperature.The decay rate $\gamma = \Gamma/H_0$ governs the $s \rightarrow \nu$ decays of steriles into active neutrinos which add to the radiation density of the Universe, and is given by:
\begin{equation}
\Gamma _ { s } = \frac { G _ { F } ^ { 2 } } { 192 \pi ^ { 3 } } m _ { s } ^ { 5 } \sin ^ { 2 } \theta.
\end{equation}

\begin{figure}
\centering
\includegraphics[width=0.5\textwidth]{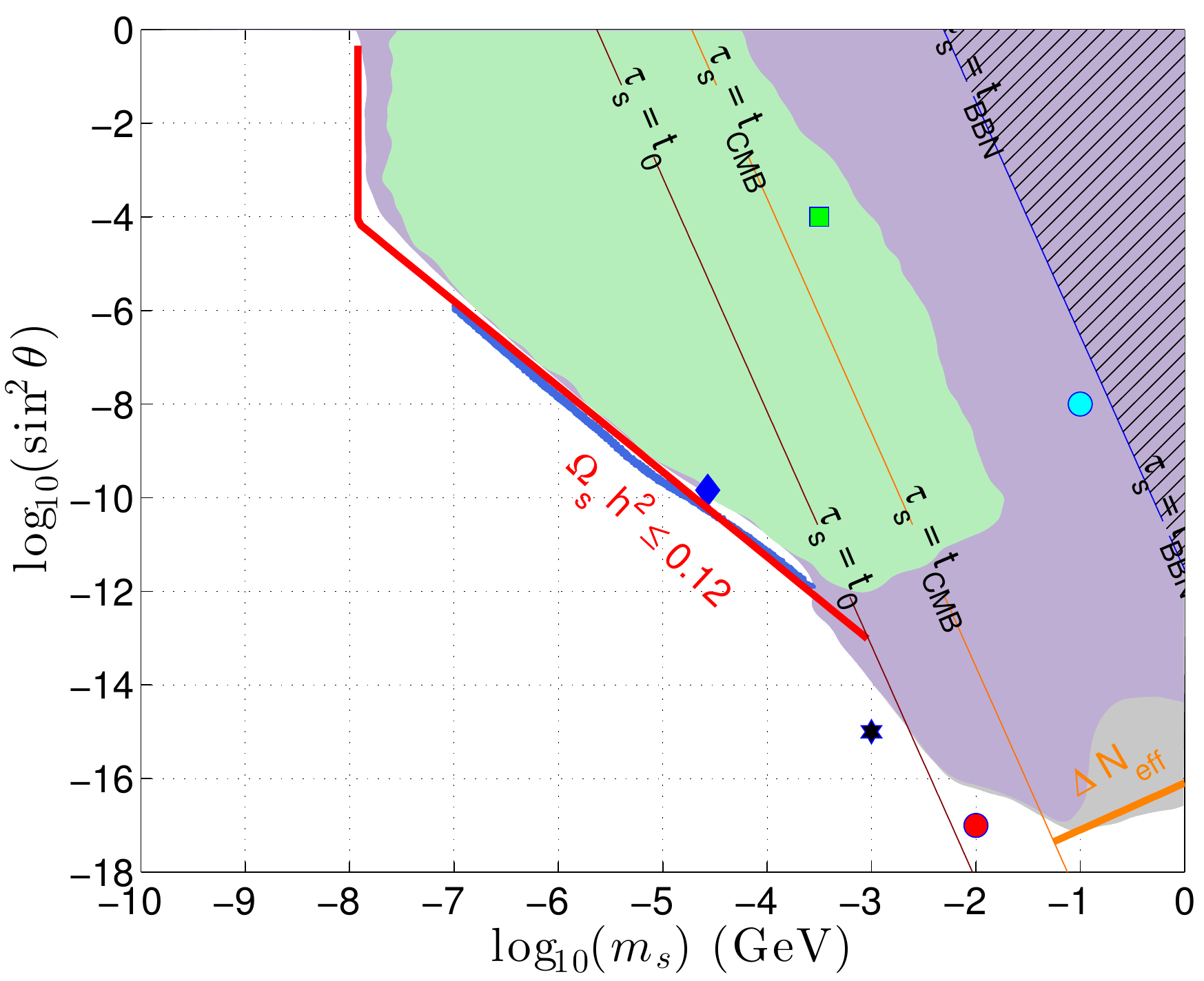}
\caption{ Constraints on a new sterile neutrino species based on freeze-in cosmological production. Coloured regions are disallowed.  Light green region: $H_0$ only. Purple region: full data set including BAO. Particles that decay before BBN (hashed region) cannot be constrained with this method.  Red and orange lines respectively correspond to estimated limits on the total abundance of DM $\Omega \lesssim0.12$, and to the requirement that $\Delta_{Neff} \lesssim 1$. The darker blue line represents the region for which the sterile neutrino can act as the cosmological dark matter. Finally, the thin lines are lines of equal sterile neutrino lifetimes, corresponding to decays that occur at present ($t_0$ = 13.5 Gyr), at recombination ($t_{CMB}$ = 380 000 yr) and at BBN ($t_{BBN}$ = 10 s). Figure from Ref. \cite{Vincent:2014rja}.}
\label{fig:nuhubble}
\end{figure}

By requiring that the Hubble rate today, $H_0$, satisfy observations, one can already rule out a large swath of parameter space which leads to an overclosed universe. This disallowed region is shown in green in Fig.~\ref{fig:nuhubble}. Note that we allow for an extra component of dark matter if it is not already saturated by the sterile neutrino contribution. By adding information on the cosmological expansion rate at different epochs, one can very quickly expand such an analysis. The purple region of Fig. \ref{fig:nuhubble} becomes disallowed by adding Baryon Acoustic Oscillation (BAO) data from WiggleZ \cite{2011MNRAS.418.1707B}, BOSS DR11 \cite{2013AJ....145...10D,Font-Ribera:2013wce} and SDSSIII \cite{2011AJ....142...72E}.

\section{Example 2: Integral and the 511 keV line}
\label{sec:511}
For over four decades \cite{1978ApJ...225L..11L}, a strong 511 keV gamma ray line has been seen from a spherically symmetric region in the inner 10$^\circ$ of the Milky Way galaxy. This flux of $\sim 10^{-3}$ photons cm$^{-2}$ s$^{-1}$ corresponds to a steady-state annihilation rate of $\sim 10^{43}$ $e^+e^-$ pairs per second. The origin of this signal remains unknown. A particular source of consternation is the morphology of this signal: even though there is a fainter disk component that can be accounted for by radioactive $\beta^+$ sources such as $^{26}Al$, $^{44}$Ti and $^{56}$Ni, the large bulge component seen most recently by INTEGRAL/SPI \cite{Siegert:2015knp} does not trace any known distribution of astrophysical sources.

It was pointed out \cite{Boehm:2003bt} that an annihilation signal from light WIMP dark matter would produce a morphology that is very similar to the one observed by INTEGRAL. A DM particle annihilating to $e^\pm$ pairs would need to be light enough ($\lesssim 10$ MeV) to avoid overproducing gamma rays via final-state radiation \cite{Beacom:2005qv}. The required cross section is far below the thermal relic cross section: $\langle \sigma v \rangle \sim 10^{-31} (m_\chi/\mathrm{MeV})^2$ cm$^3$ s$^{-1}$ \cite{Vincent:2012an}. This is not a problem, as long as another channel exists to avoid overproduction of DM in the early universe. In \cite{Wilkinson:2016gsy}, we explored the possible ways of consistently obtaining the relic abundance by freeze-out: either with an extra p-wave component that is suppressed at late times ($\langle \sigma v \rangle = a + b v^2$, where $a$ is responsible for the galactic centre excess and $b \gg a$ sets the relic density), or by annihilation mainly into the neutrino sectors. Both of these options can affect early Universe cosmology. 

As DM becomes non-relativistic, its entropy is dumped into the remaining relativistic degrees of freedom. If this occurs after neutrino have decoupled from the thermal bath, this can change the relative entropy between the electromagnetic and neutrino sectors, leading in turn to a larger (or smaller) radiation content than would be expected in the standard scenario. This is parametrized with the quantity $\Delta N_{eff}$, which represents the amount of extra entropy in the neutrino sector. If the DM decouples from the electromagnetic sector (i.e. has an effective coupling to $e^\pm$), then this reheating of the visible sectors leads to a negative value of $\Delta N_{eff}$ since the neutrino sector is now colder by comparison \cite{Boehm:2012gr}. This quantity directly affects the expansion rate of the Universe, and is thus measurable by its effect on the abundances of primordial elements (faster expansion causes neutron abundance to be less Boltzmann-suppressed at freeze-out, thus leading to more primordial helium and deuterium \cite{Sarkar:1995dd}) as well as the acoustic peaks of the CMB. DM-neutrino scattering also leads to damping of the matter power spectrum at small scales (``collisional damping'' \cite{Boehm:2000gq}).

Finally, the small s-wave annihilation rate to $e^\pm$, responsible for the 511 keV signal, would also contribute to ionizing the intergalactic medium during the dark ages (we refer to this as \textit{raising the ionization floor}. This extra free electron population rescatters CMB photons as they propagate the 13.6 billion lightyears between the last scattering surface and present-day Earth. This leads to significant effects on the temperature and polarization anisotropy spectra of the CMB \cite{Chen:2003gz}. 

\begin{figure}
\centering
\includegraphics[width=0.35\textwidth,angle=-90,clip,trim=0ex 0ex 0ex 5ex]{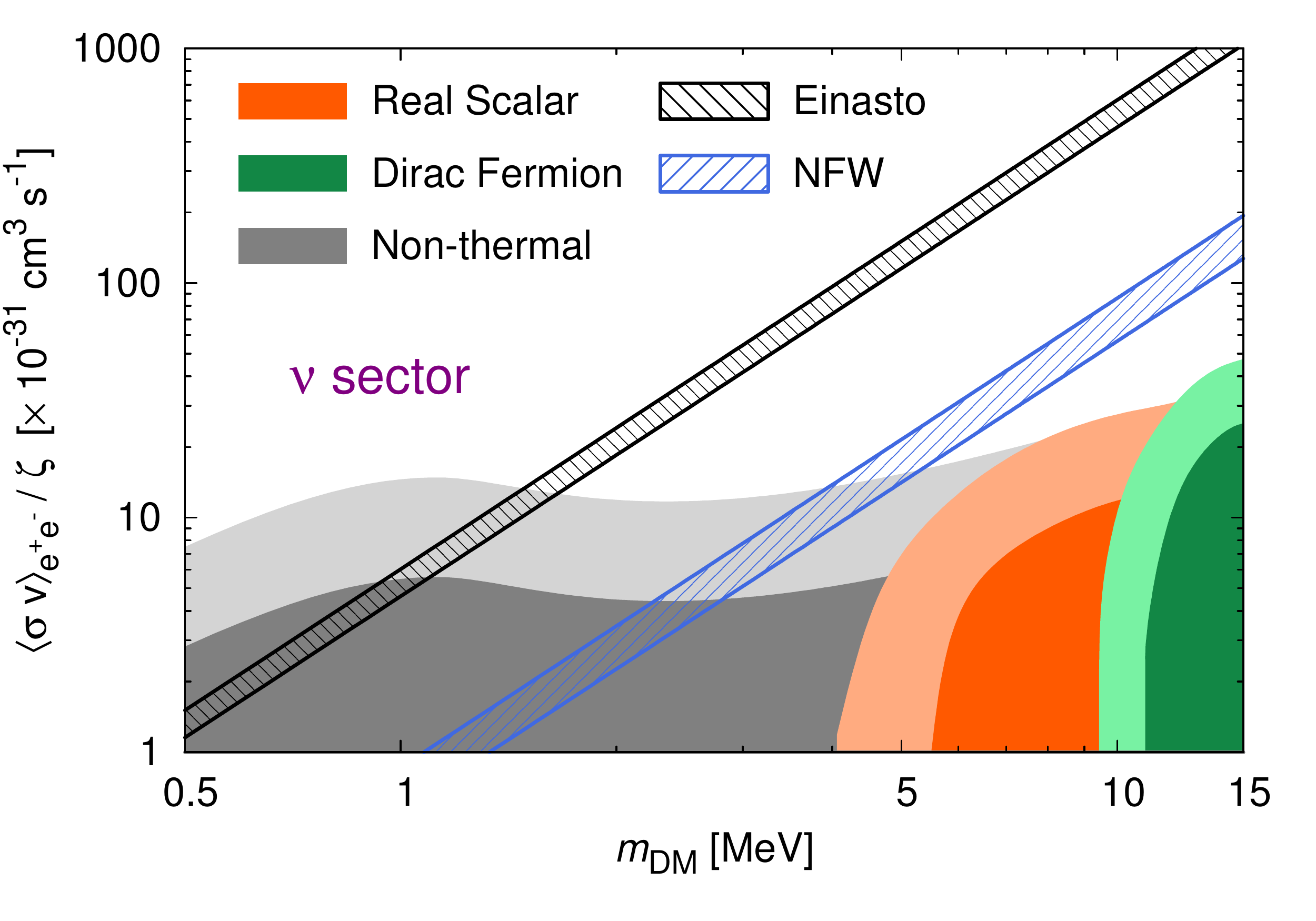}
\hspace{1ex}
\includegraphics[width=0.35\textwidth,angle=-90,clip,trim=0ex 0ex 0ex 5ex]{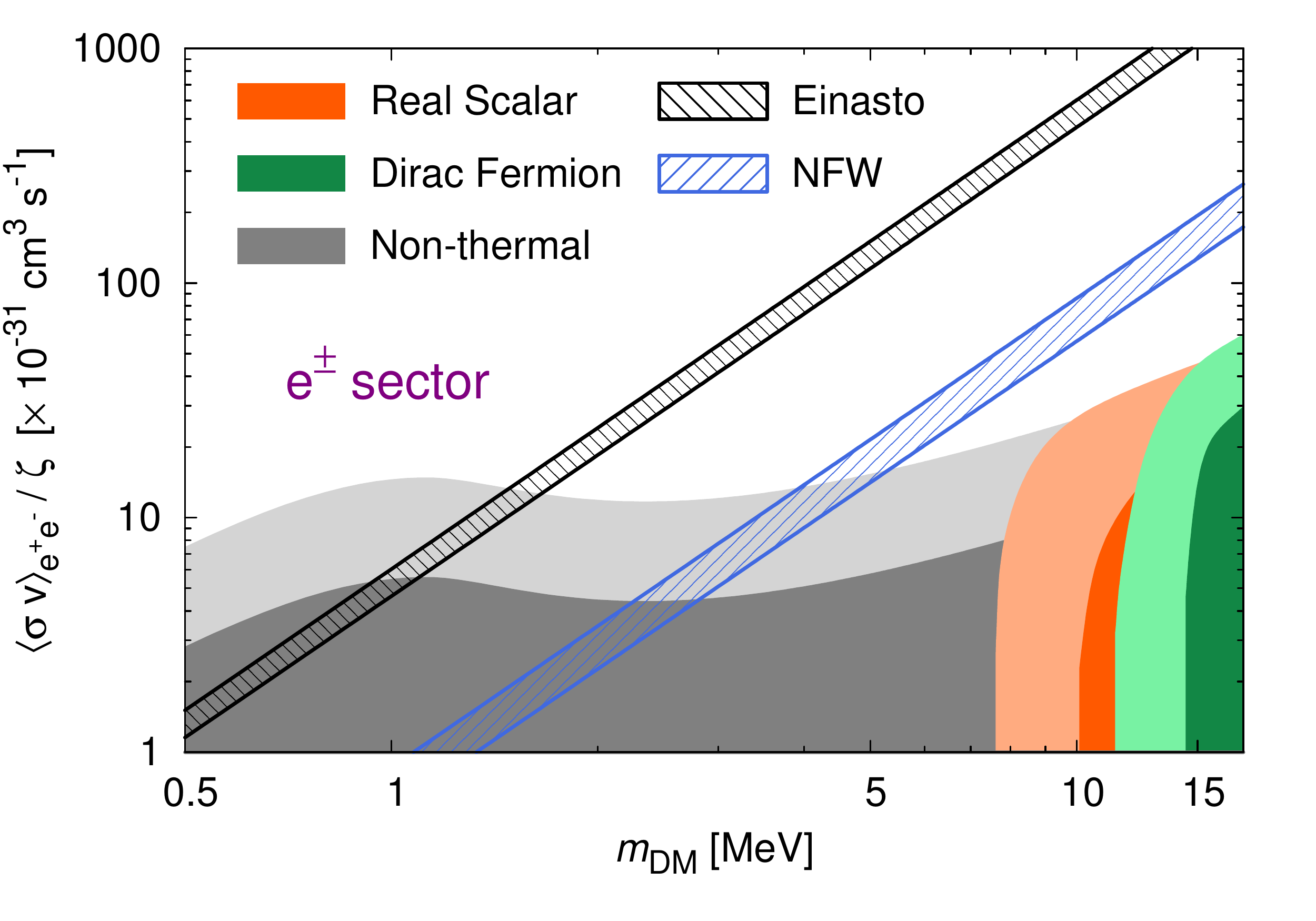}
\vspace{-2ex}
%\end{center}
\caption{Cosmological limits due to the ionization of the intergalactic medium (grey) expansion of the Universe (colours) strongly constrain scenarios in which a thermal WIMP produces the observed galactic 511 keV excess. Note that coloured regions are \textit{allowed}. Figures from \cite{Wilkinson:2016gsy}.}
\label{fig:ryan}
\end{figure}

When the smoke clears, one finds that scenarios that correctly produce a WIMP via freeze-out, which can plausibly explain the INTEGRAL/SPI 511 keV excess, appear to be disallowed by cosmology. Fig.~\ref{fig:ryan}, taken from Ref.~\cite{Wilkinson:2016gsy}, shows these constraints. The textured lines show the required combination of mass and cross section to explain the INTEGRAL signal. The grey bands show what is allowed (1 and 2 $\sigma$) from the ionization floor. Finally, the coloured sections show the regions that are allowed by a full Monte Carlo parameter space search, which includes constraints from $\Delta N_{eff}$, the ionization floor and collisional damping. 
The 511 keV story is thus far from resolved: adding to the mystery is the recent observation of a signal from at least one dwarf spheroidal galaxy, Reticulum II \cite{Siegert:2016ijv}, the same dwarf galaxy that may also harbour a $\sim$ GeV energy gamma ray signal \cite{Geringer-Sameth:2015lua}. 

\section{Conclusions}
\label{sec:conclusions}
I have provided two compelling examples illustrating the interplay between particle physics, astrophysics and cosmology. The Universe is a laboratory for particle physics. As long as  backgrounds can be correctly quantified, astrophysical and cosmological observations have the power to provide tremendous input to our knowledge of particle physics. Several of the elementary particles, including the positron and the muon, were discovered via their presence in cosmic rays and much of what we know of nuclear physics and neutrinos is thanks to astrophysical processes that occurred far beyond Earth. Looking forward, with the myriad new observatories, space telescopes and surveys, the aspiring particle physicist will be wise to keep their gaze pointed towards the stars.

\section*{Acknowledgements}
I thank P. Pietroff for the invitation to EDSU, as well as the organizing committee and local government for a wonderfully productive workshop. I acknowledge support from the Arthur B. McDonald Institute, which is made possible by the Canada First Research Excellence Fund (CFREF).

\bibliographystyle{JHEPmod}
\bibliography{G18,511planck}

\end{document}